\documentclass[sigconf, screen]{acmart}

\usepackage{multirow}
\usepackage{subcaption}

\usepackage{float}
\AtBeginDocument{%
  }

\setcopyright{none} 

\copyrightyear{2026}
\acmYear{2026}
\acmConference[FSE '26]{Companion Proceedings of the 34th ACM Symposium on the Foundations of Software Engineering}{June 5--9, 2026}{Montreal, Canada}
\acmBooktitle{Companion Proceedings of the 34th ACM Symposium on the Foundations of Software Engineering (FSE '26), June 5--9, 2026, Montreal, Canada}

\acmISBN{}

\begin{document}

\title{SERSEM: Selective Entropy-Weighted Scoring for Membership Inference in Code Language Models}


\author{Kıvanç Kuzey Dikici}
\authornote{Equal contribution with alphabetical order.}
\affiliation{%
  \institution{Bilkent University}
  \department{Computer Engineering}
  \city{Ankara}
  \country{Turkey}}
\email{kuzey.dikici@ug.bilkent.edu.tr}
\orcid{0009-0001-1709-9102}

\author{Serdar Kara}
\authornotemark[1]
\affiliation{%
  \institution{Bilkent University}
  \department{Computer Engineering}
  \city{Ankara}
  \country{Turkey}}
\email{serdar.kara@ug.bilkent.edu.tr}
\orcid{0009-0005-6238-0963}

\author{Semih Çağlar}
\affiliation{%
  \institution{Bilkent University}
  \department{Computer Engineering}
  \city{Ankara}
  \country{Turkey}}
\email{semih.caglar@ug.bilkent.edu.tr}
\orcid{0009-0003-1311-5747}

\author{Eray Tüzün}
\affiliation{%
  \institution{Bilkent University}
  \department{Computer Engineering}
  \city{Ankara}
  \country{Turkey}}
\email{eraytuzun@cs.bilkent.edu.tr}
\orcid{0000-0002-5550-7816}

\author{Sinem Sav}
\affiliation{%
  \institution{Bilkent University}
  \department{Computer Engineering}
  \city{Ankara}
  \country{Turkey}}
\email{sinem.sav@cs.bilkent.edu.tr}
\orcid{0000-0001-9096-8768}

\renewcommand{\shortauthors}{Dikici et al.}

\begin{abstract}
As Large Language Models (LLMs) for code increasingly utilize massive, often non-permissively licensed datasets, evaluating data contamination through Membership Inference Attacks (MIAs) has become critical. We propose \textbf{SERSEM} (Selective Entropy-Weighted Scoring for Membership Inference), a novel white-box attack framework that suppresses uninformative syntactical boilerplate to amplify specific memorization signals. SERSEM utilizes a dual-signal methodology: first, a continuous character-level weight mask is derived through static Abstract Syntax Tree (AST) analysis, spellchecking-based multilingual logic detection, and offline linting. Second, these heuristic weights are used to pool internal transformer activations and calibrate token-level Z-scores from the output logits. Evaluated on a 25,000-sample balanced dataset, SERSEM achieves a global \textbf{AUC-ROC of 0.7913} on the StarCoder2-3B model and \textbf{0.7867} on the StarCoder2-7B model, consistently outperforming the implemented probability-based baselines Loss, Min-K\% Prob, and PAC. Our findings demonstrate that focusing on human-centric coding anomalies provides a significantly more robust indicator of verbatim memorization than sequence-level probability averages.
\end{abstract}




\maketitle

\section{Introduction}

Large Language Models (LLMs) trained on source code have demonstrated remarkable capabilities in code generation, completion, and reasoning \cite{lozhkov2024starcoder2}. Nonetheless, since these models scale in size and are exposed to massive corpora such as The Stack v2 \cite{kocetkov2022stack}, they increasingly exhibit a tendency to memorize portions of their training data \cite{alkaswan2024traces}. This memorization poses significant legal and ethical risks, particularly when models inadvertently reproduce non-permissively licensed or proprietary code snippets such as those curated in The Heap dataset \cite{theheap}. 

To assess the extent of data contamination in closed-data models, researchers rely on Membership Inference Attacks (MIAs) \cite{shokri2017membership}. In a general white-box membership inference setting, the adversary attempts to determine whether a specific data point was included in the target model's training set by computing probabilities over the target model \cite{carlini2022quantifying, shokri2017membership, li2024unveilingunseenexploringwhitebox}. Although standard baselines such as negative log-likelihood (Loss), Min-K \% Prob \cite{shi2023detecting}, Surprise-based MIAs (SURP) \cite{zhang2024surp}, and Polarized Augment Calibration (PAC) \cite{maini2024pac} have been proposed, they fundamentally treat source code as a natural language. Source code, however, is structurally bound by rigid syntax (e.g., \texttt{for}, \texttt{def}, \texttt{return}). Because an LLM can predict standard boilerplate with near-perfect confidence regardless of whether it has memorized the specific file, these probability techniques suffer from high false-positive rates when applied to software artifacts \cite{zhang2024surp, shi2023detecting}.

The Poisoned Chalice Competition 2026 \cite{katzy2026poisoned} establishes a white-box evaluation framework to advance domain-specific MIAs against code models. In this paper, we present our solution: \textbf{SERSEM} (Selective Entropy-Weighted Scoring for Membership Inference). SERSEM shifts the attack vector from general sequence prediction to continuous targeted anomaly detection. We assert that prediction of human fallibility, such as formatting inconsistencies, emotional comments, multilingual logic slippage, and idiosyncratic naming conventions, can provide a useful signal of verbatim memorization.

Furthermore, our approach aligns the heuristic coding weights with deep internal activation profiles. By extracting membership signals from the intermediate layers of the transformer rather than relying exclusively on final output logits, we isolate the representational signatures of memorization before the model optimizes for syntactical correctness \cite{ibanezlissen2024lumia}. By continuously weighting these internal and output probability distributions based on the entropic density of the code segment, our approach suppresses boilerplate noise and improves MIA accuracy against data-blurred code models.

\section{Methodology}

\begin{figure*}[t]
    \centering
    \includegraphics[width=\textwidth]{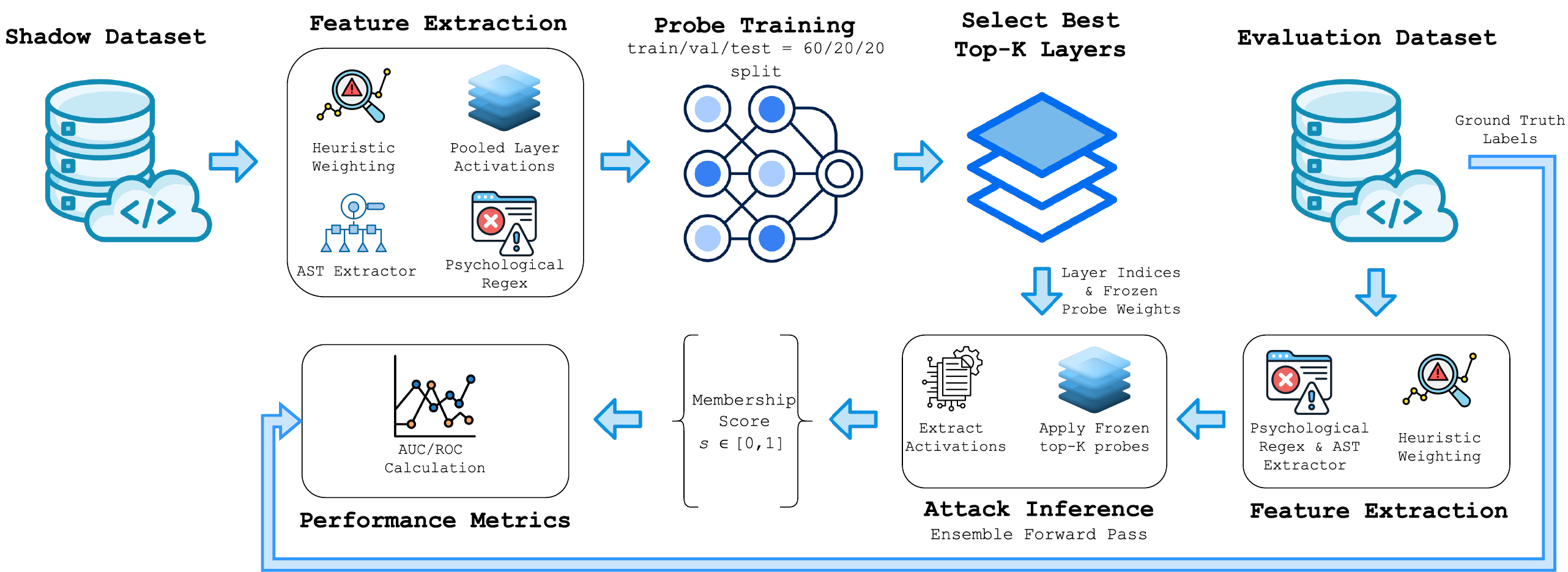}
    \caption{Architecture and workflow of the SERSEM framework.}
    \label{fig:methodology}
\end{figure*}

This section details the design and implementation of the SERSEM framework, an attack specifically engineered to exploit the fine-grained distinctions of memorized source code. The core architecture of SERSEM utilizes a dual-angled approach (see Figure \ref{fig:methodology}). First, it extracts human-centric anomalies through static code analysis to construct a continuous character-level weight mask. Second, it utilizes both the final output logits and the deeply embedded transformer activations to derive a final probability score.

\subsection{Continuous Entropy-Weighted Scoring}
\label{subsec:scoring}
Standard source code comprises highly predictable repeating boilerplate syntax. When these tokens are generated, standard causal LMs yield extremely high probabilities regardless of memorization, making the genuine membership signal thinner \cite{shi2023detecting, zhang2024surp}. 

To counteract this, \texttt{WeightedAnomalyMIA} assigns a specific continuous weight to each token in the evaluation sequence, prioritizing tokens that reflect human irregularity. We construct a custom anomaly map by extracting six critical continuous signals via an Abstract Syntax Tree (AST) using \texttt{tree-sitter} \cite{treesitter} and secondary heuristics:

\begin{itemize}
    \item \textbf{Boilerplate Noise Mitigation}: To suppress the noise of predictable syntax and punctuation, tokens failing a structural identifier regex match (e.g., brackets, operators, keywords, short auxiliary variables) are penalized and downgraded to a negligible weight.
    \item \textbf{Standard Identifiers}: Code tokens that strictly match valid programmable variable formatting are preserved at a baseline standard weight.
    \item \textbf{Long Identifiers}: To specifically prioritize developer-centric naming, identifiers exceeding 10 characters are assigned an elevated weight by the AST.
    \item \textbf{Literal Strings}: AST parsing dynamically inflates weights for high-entropy data artifacts such as string literals.
    \item \textbf{Linter Formatting Errors}: Offline linting maps abnormal spacing, inconsistent indentation, and arbitrary structural chaos to absolute character offsets. For Python, this is executed via \texttt{flake8} \cite{flake8}; other languages utilize generic regex fallbacks.
    \item \textbf{Comments \& Multilingual Slippage}: Standard comments mapped via the AST, and identifiers failing an \texttt{en\_US} \texttt{PyEnchant} \cite{pyenchant} dictionary check, receive a severe penalty for demonstrating out-of-distribution human blunders.
    \item \textbf{Psychological Tags}: An override regex specifically targets high-confidence developer signatures (e.g., \texttt{TODO}, \texttt{FIXME}).
\end{itemize}

During inference, these character-level heuristic signals are projected onto the subword tokens generated by the tokenizer. The final consolidated token weights that drive the anomaly scoring mechanism are summarized in Table~\ref{tab:weights}.

\begin{table}[h]
\centering
\caption{SERSEM Final Token Weighting Scheme.}
\label{tab:weights}
\begin{tabular}{lc}
\hline
\textbf{Token Classification} & \textbf{Weight} \\ \hline
Boilerplate Noise \& Syntax & 0.1 \\
Standard Identifiers ($<10$ chars) & 1.0 \\
Long Identifiers ($\ge10$ chars) & 3.0 \\
Literal Strings & 5.0 \\
Linter Formatting Errors & 5.0 \\
Comments \& Multilingual Slippage & 10.0 \\
Psychological Tags (\texttt{TODO}) & 10.0 \\ \hline
\end{tabular}
\end{table}

Instead of evaluating raw output probabilities, SERSEM computes the \textbf{Z-Score} of the correct next token's logit relative to the full vocabulary distribution at that prediction step. By passing the Z-Scores through a sigmoid layer and multiplying them against the normalized token weight array, the final score inherently grades the LLM exclusively on how good it predicts human blind spots while suppressing structural boilerplate.

\subsection{Internal Activation Probing }
Although the output logits provide significant predictive insight, the final layers of a transformer are biased towards syntactical generation \cite{ibanezlissen2024lumia}. Verbatim memorization signatures are typically strongest within the hidden states of intermediate layers. To capture these deep representational signals, we integrate LUMIA (Linear probing for Unimodal and MultiModal Membership Inference Attacks leveraging internal LLM states) \cite{ibanezlissen2024lumia}. The core assertion of LUMIA is that the internal parameter activations inside a white-box language model preserve distinguishable characteristics of its training data before they are smoothed out by the final classification head.

Leveraging the white-box constraints of the Poisoned Chalice Competition \cite{katzy2026poisoned}, SERSEM attaches dynamic forward hooks to every transformer layer of the target model. We extract the internal activations during a single, no-gradient forward pass. Because pure mean-pooling risks flattening the signal of rare keywords, we independently pool the activations using both standard averaging and the entropy-based structural weights described in Section~\ref{subsec:scoring}. These representations are concatenated to yield an augmented activation space.

Following the LUMIA methodology \cite{ibanezlissen2024lumia}, we evaluate a random subset of validation shadow data to extract layer activations for known members and non-members. These extracted features allow us to dynamically select the most discriminative hidden layers. A lightweight Multi-Layer Perceptron (MLP) probe is trained via binary cross-entropy on the concatenated internal representations for each chosen layer to predict binary membership. In inference, the attack dynamically ensembles the sigmoid outputs from the top-performing five layer probes, outputting a precise membership 230 confidence score bounded between [0.0, 1.0].

\section{Experiments and Results}
This section presents the empirical evaluation of SERSEM, covering the experimental setup, the performance of the per-language membership inference, and a discussion of the detection variability across programming paradigms and baseline comparisons. The AUC-ROC curves for both target models are illustrated in Figure~\ref{fig:roc_curves}.

\subsection{Experimental Setup}
To conform to the competition structure while addressing the lack of a ground-truth held-out test set during validation, we partitioned the provided Poisoned Chalice development dataset \cite{katzy2026poisoned}. For each of the five target languages (Python, Java, Go, Ruby, Rust), we extracted a random sampling of 10,000 instances \textbf{balanced equally} between members (The Stack v2 \cite{lozhkov2024starcoder2}) and non-members (The Heap \cite{theheap}).

From this pool, 5,000 samples per language (25,000 total) were allocated for training our intermediate layer probe-training \footnote{The trained shadow model weights are publicly available on HuggingFace: \url{https://huggingface.co/Serdark4r/SERSEM-Poisoned-Chalice-Competition-2026}}. To ensure the complete integrity of our Results section and the final competition submission, we enforce two separate layers of data isolation. First, we maintain a no-overlap property between our internal training and inference splits (5,000 samples each per language). Second, we have confirmed with the organizers that the official held-out evaluation set used for final scoring has zero intersection with the released development data. 

The SERSEM pipeline was developed and validated on a high-performance workstation featuring a single NVIDIA TITAN RTX (24GB VRAM) GPU, dual Intel(R) Xeon(R) Gold 6140 CPUs (36 physical cores, 72 threads), 256GB of system RAM, and a 2.2TB high-speed data partition. We evaluated our approach on the \textbf{StarCoder2-3B} and \textbf{StarCoder2-7B} models \cite{lozhkov2024starcoder2}, which were pre-trained on \textbf{The Stack v2} dataset \cite{lozhkov2024starcoder2}. This ensures that samples drawn from The Stack v2 are verifiably members of the model's training corpus. Due to the architectural limitations of our local validation environment (TITAN RTX), all experiments reported in this paper were conducted utilizing \texttt{float16} precision and sequence lengths of up to 8,196 tokens. However, our proposed submission pipeline is fully compatible with \texttt{bfloat16} precision for the final evaluation on the competition's NVIDIA A100 hardware, ensuring optimal performance within the 12-hour wall-clock limit.

\subsection{Empirical Analysis of Detection Accuracies}
\begin{figure}[t]
    \centering
    \begin{subfigure}[b]{0.48\columnwidth}
        \centering
        \includegraphics[width=\textwidth]{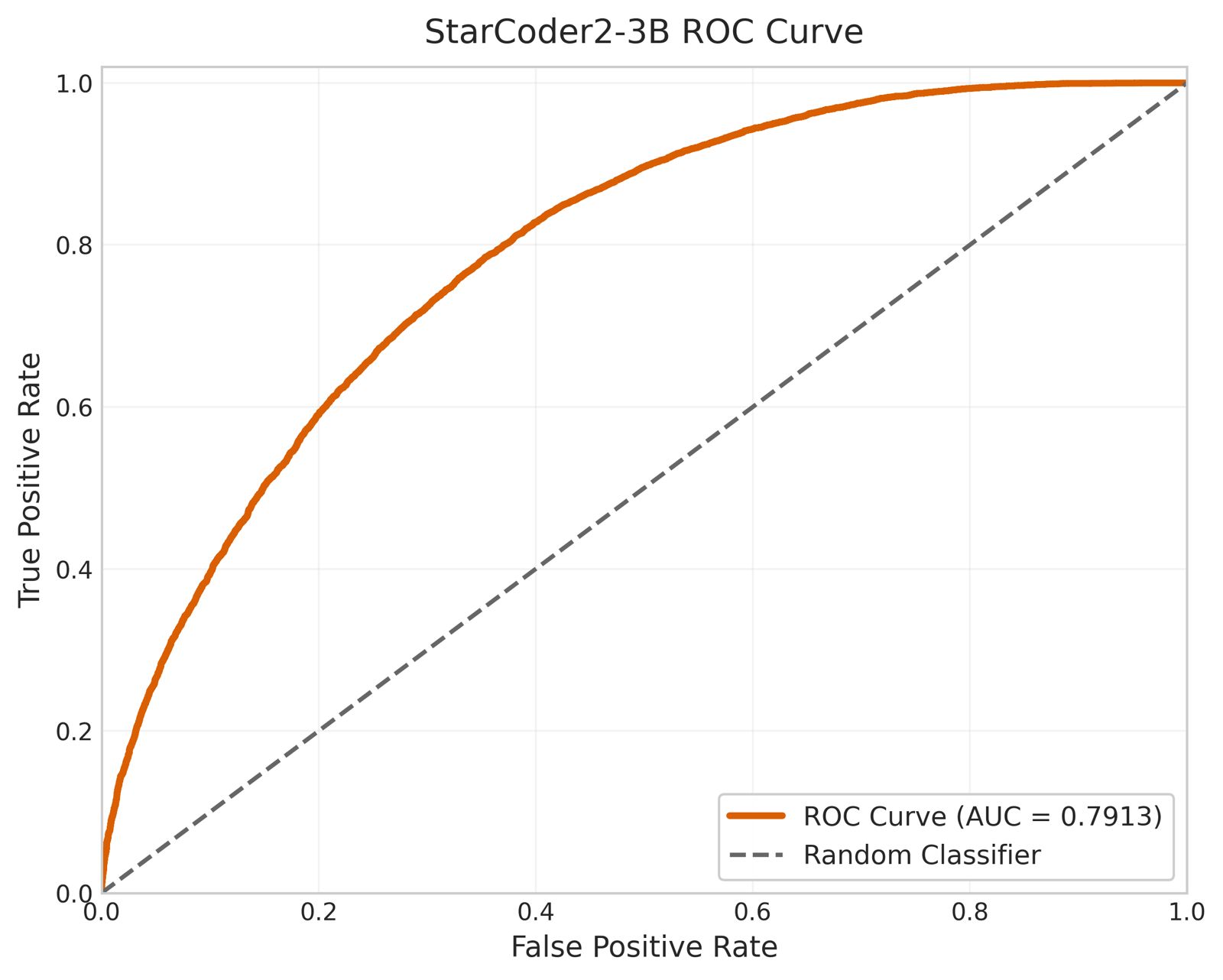}
        \caption{StarCoder2-3B}
        \label{fig:roc_3b}
    \end{subfigure}
    \hfill
    \begin{subfigure}[b]{0.48\columnwidth}
        \centering
        \includegraphics[width=\textwidth]{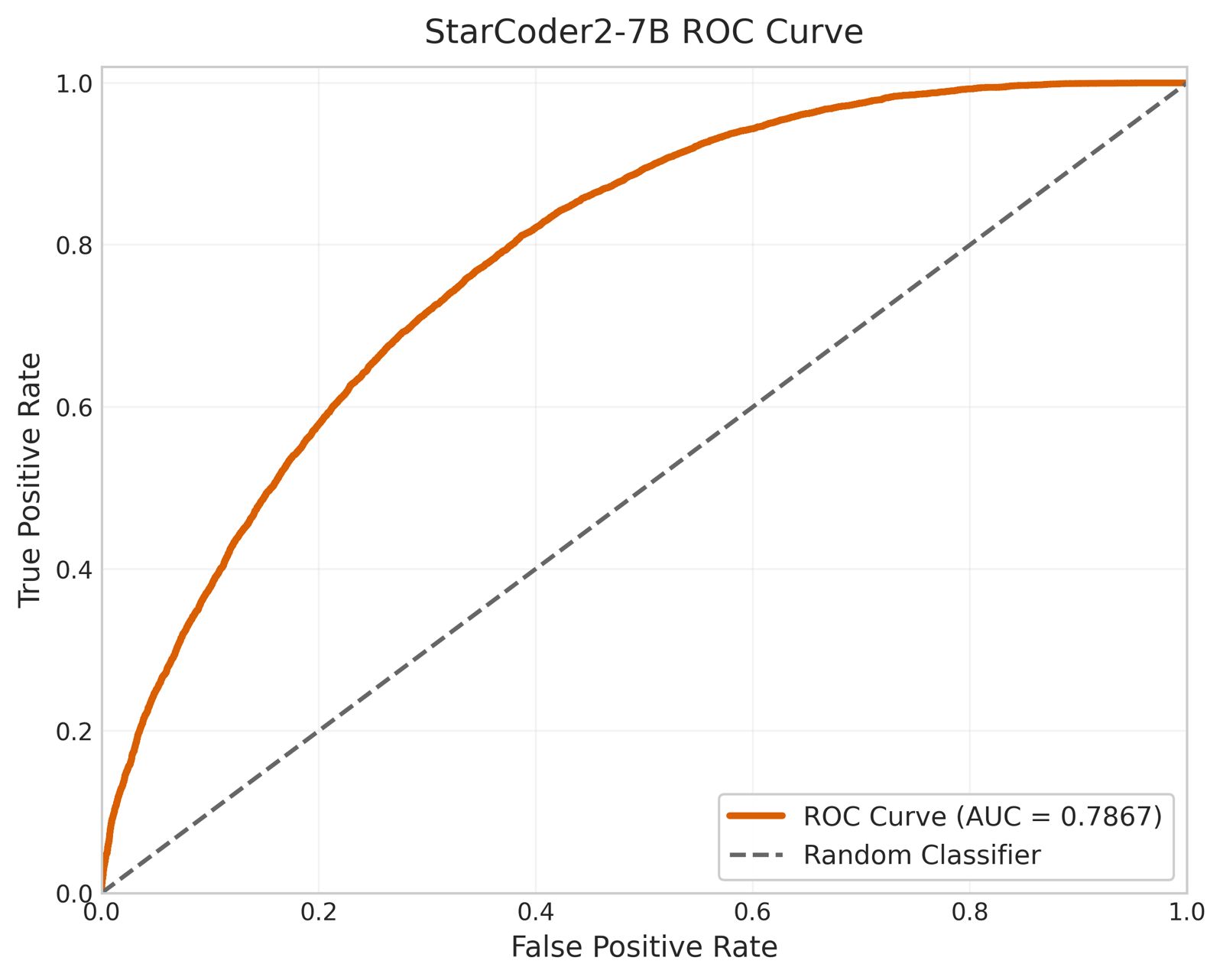}
        \caption{StarCoder2-7B}
        \label{fig:roc_7b}
    \end{subfigure}
    \caption{SERSEM AUC-ROC curves.}
    \label{fig:roc_curves}
\end{figure}

Evaluating our full pipeline on held-out inference splits of 5,000 samples per language, SERSEM achieves an overall AUC-ROC of \textbf{0.7913} on StarCoder2-3B and \textbf{0.7867} on StarCoder2-7B (Figure~\ref{fig:roc_curves}), indicating strong membership detection performance. The Loss, Min-K\% Prob, and PAC baselines in Table~\ref{tab:lang_results} were run using the official PoisonedChalice implementations\footnote{\url{https://github.com/AISE-TUDelft/PoisonedChalice}}; due to local hardware constraints, PAC was executed in \texttt{float16} rather than \texttt{bfloat16}, consistent with our local validation setup. 

To provide granular insight into the attack's effectiveness across different syntax structures, a detailed breakdown of the AUC-ROC performance per programming language is presented in Table~\ref{tab:lang_results}.

\begin{table}[t]
\centering
\caption{Per-language AUC-ROC comparison of SERSEM and implemented baselines on 5,000 samples per language.}
\label{tab:lang_results}
\resizebox{\columnwidth}{!}{%
\begin{tabular}{lccccc}
\hline
\textbf{Language} & \textbf{SERSEM 3B} & \textbf{SERSEM 7B} & \textbf{Loss 3B} & \textbf{MKP 3B} & \textbf{PAC 3B} \\ \hline
\textbf{Overall}  & \textbf{0.7913} & \textbf{0.7867} & \textbf{0.5881} & \textbf{0.5931} & \textbf{0.4712} \\ \hline
Go     & 0.8121 & 0.8071 & 0.5978 & 0.6016 & 0.5023 \\
Java   & 0.7738 & 0.7632 & 0.5856 & 0.5906 & 0.4637 \\
Python & 0.8066 & 0.8045 & 0.6027 & 0.6071 & 0.4700 \\
Ruby   & 0.8155 & 0.8151 & 0.6302 & 0.6367 & 0.4237 \\
Rust   & 0.7402 & 0.7379 & 0.5643 & 0.5658 & 0.4898 \\ \hline
\end{tabular}}
\end{table}

\subsection{Discussion of Language Variability}
The results highlight a significant divergence in detection performance between programming paradigms. Ruby (0.8155) and Go (0.8121) achieved the highest AUC scores for the 3B model, while Ruby (0.8151) remained the language that performed the best for the 7B variant. These high scores are likely due to the flexible, highly idiosyncratic formatting rules and human conventions in these languages, which allow for greater expression of human coding chaos. Python also exhibited consistently strong performance across both model sizes (0.8066 for 3B and 0.8045 for 7B), suggesting that although Python has a more standardized and readable syntax than Ruby, it still preserves substantial room for human-specific variation through comments, naming styles, spacing behavior, and informal scripting conventions. In this sense, Python appears to occupy a middle ground: it is less structurally rigid than Java and Rust, but less stylistically unconstrained than Ruby, making it particularly suitable for SERSEM’s anomaly-focused weighting strategy. Conversely, Rust (0.7402) and Java (0.7738) enforce stricter structural and typed boilerplate, which inherently suppress unique human variations and reduce the impact of the static continuous penalty mask. 

\subsection{Discussion of Baseline Comparisons}
As shown in Table~\ref{tab:lang_results}, SERSEM consistently outperforms all three baselines implemented on the StarCoder2-3B model. Loss, Min-K\% Prob (MKP), and PAC achieve overall AUC-ROC values of 0.5881, 0.5931, and 0.4712, respectively, whereas SERSEM reaches 0.7913, representing an improvement of more than 20 percentage points. This margin is consistent across all five languages: even in Rust, where SERSEM attains its lowest AUC (0.7402), it remains substantially above the strongest baseline. This performance gap reflects a limitation shared by all three baselines: they rely exclusively on output-probability information and remain agnostic to the structural role each token plays in source code. Loss aggregates token-level likelihood across the sequence, PAC calibrates sequence-level confidence using augmented neighboring samples, and even the more selective Min-K\% Prob focuses only on the least likely tokens without incorporating code structure. As a result, these methods cannot distinguish between tokens that are difficult because they reflect genuine training data exposure and tokens that are simply rare under the model's general distribution. SERSEM addresses this limitation by grounding its weighting scheme in code structure.

\section{Conclusion}
In this paper, we present \textbf{SERSEM}, a white-box membership inference attack designed for code language models. By shifting from global probability distributions to structure-aware anomaly detection, SERSEM isolates human-centric memorization signals in source code. Our method combines static AST-based heuristics and automated linting with internal activation probes, allowing it to leverage deeper representational signals that are often obscured in the final output logits. Evaluated on a 25,000-sample dataset, SERSEM achieved a robust AUC-ROC of \textbf{0.7913} against StarCoder2-3B and \textbf{0.7867} against StarCoder2-7B. It also consistently outperforms the implemented probability-based baselines Loss, Min-K\% Prob, and PAC on our held-out inference split. Future work includes extending entropy-weighted masks to broader language paradigms, automatically tuning heuristic weights per target distribution, and replacing regex-based fallbacks with native linters for non-Python languages. This paper is submitted as part of the \textit{Poisoned Chalice Competition} \cite{katzy2026poisoned}, and the replication package for this work is publicly available\footnote{\url{https://doi.org/10.6084/m9.figshare.31911933}}.


\bibliographystyle{ACM-Reference-Format}
\bibliography{references}

\end{document}